\documentclass{article}
\usepackage{graphicx}

\def\href#1#2{\texttt{#2}}

\def\vrulefill{\leaders\vrule\vfill}

\newcommand{\page}[3]{%
  \hbox to\hsize{%
    \scriptsize
    \hfill
    \hbox{%
      \vbox to3cm{%
        \vrulefill
      }%
      \vbox to3cm{%
        \hbox to6cm{\hrulefill}
        \medskip
        \hbox to6cm{\hspace{4pt}#1\hspace{4pt}}
        \medskip
        \hbox to6cm{\hspace{4pt}#2\hfill\hspace{4pt}}
        \hbox to6cm{\hspace{4pt}#3\hfill\hspace{4pt}}
        \vfill
      }%
      \vbox to3cm{%
        \vrulefill
      }
    }%
    \hfill
  }
}
\title{Quasi 2D Bose-Einstein condensation\\ in an optical lattice}
\author{S.~Burger, F.\,S.~Cataliotti, C.~Fort, P.~Maddaloni,\\  F.~Minardi, and
 M.~Inguscio\\
Laboratorio Europeo di Spettroscopia Nonlineare (LENS),\\
Istituto Nazionale per la Fisica della Materia,\\
Dipartimento di Fisica dell' Universit\`a di Firenze,\\ 
Largo~Enrico~Fermi,~2, I-50125 Firenze, Italia}

\begin{document}

\maketitle

\begin{abstract}
We study the phase transition of a gas of $^{87}$Rb atoms to quantum degeneracy
in the combined potential of a harmonically confining magnetic
trap and the periodic  potential of an optical lattice.
For high optical lattice potentials 
we observe a significant change in the 
temperature dependency of the population of the ground state of the system.
The experimental results are in good agreement with a model assuming
the subsequent formation
of quasi 2D condensates in the single lattice sites.
\end{abstract}


\vspace{4pc}

The physical behaviour of a system is strongly influenced by
its dimensionality. 
A well known example is the appearance of plateaus in the Hall resistance
across a twodimensional (2D) electron gas as a function of the number of 
electrons
(quantum Hall effect~\cite{Klitzing1986a}).

In atomic physics, important steps towards the realization of pure 2D
systems of neutral atoms have been made in different systems:
Significant fractions of atomic systems could be prepared in the 2D
potentials of optical lattices~\cite{Vuletic1998a,Bouchoule2001x} and
of an evanescent wave over a glass prism~\cite{Gauck1998a},
quasicondensates could be observed in 2D atomic hydrogen trapped
 on a surface covered with liquid $^4$He~\cite{Safonov1998a},
and 3D condensates of $^{23}$Na with low atom numbers 
could be transferred to the 2D
regime by an adiabatic deformation of the trapping 
potential~\cite{Gorlitz2001x}. 

By using Bose-Einstein condensates (BECs) confined to optical lattices it has
become possible to overcome major limitations
of previous experiments:
First, an optical lattice can confine a large array of
2D systems, which allows measurements with a much higher number
of involved atoms with respect to a single confining potential. 
Second, the macroscopic population of a single quantum state (BEC)
naturally transfers the whole system to a pure occupation of the 
2D systems, which could so far not be realized with thermal
atomic clouds.

Experiments in which BECs were
transferred~\cite{Anderson1998a,Orzel2001a,Morsch2001,Greiner2001}
or produced~\cite{Burger2001a,Cataliotti2001,Pedri2001} 
in optical lattices  concentrated on 
the measurement of ground state, tunnelling and dynamical 
properties and on atom optical applications.
The  effects on the condensation process
and the change of dimensionality  by the
superimposement of an optical lattice onto a magnetic trapping
potential have so
far not been investigated.

In this letter, we report on the Bose-Einstein condensation of a dilute gas 
of $^{87}$Rb atoms to the combined potential of a static
magnetic trap and a one-dimensional optical lattice.
Varying the temperature of the evaporatively cooled cloud 
 we probe the momentum distribution of
atomic clouds across the transition to BEC for different 
strengths of the optical lattice.
Besides a shift in the critical temperature, $T_c$,
we find a dramatic change of the temperature-dependency of the 
condensate fraction with respect to the 3D case
which can be explained by the subsequent formation of quasi 2D BECs in 
the lattice sites, each at a different transition temperature.

The dimensionality of a gas of weakly interacting bosons has important
consequences on the thermodynamic properties of the system.
While in 3D the gas undergoes the phase transition to BEC 
even in free space, in 2D systems BEC at finite temperatures can only exist
in a confining potential~\cite{Petrov2000a}.
Also, quasi-condensates, i.e., condensates with regions of uncorrelated phase
are expected to occur  in 2D, as well as in onedimensional (1D) 
and even in very elongated 3D 
systems~\cite{Petrov2000a,Petrov2001x,Dettmer2001x}.
The mechanisms leading to the lack a global condensate phase are 
quantum fluctuations and -- for a temperature $T > 0$ --
interactions with atoms from the thermal cloud. 
Similar mechanisms lead to the loss of phase coherence between 
condensates confined to different sites of
the optical lattice~\cite{Orzel2001a,Greiner2001,Pitaevskii2001x}.

For the ideal gas in a 2D harmonic trap with the fundamental frequency
$\omega$ the analytical solutions for
the condensation temperature, $T_c$, and the dependence of the condensate
fraction, $N_0/N$ (number of particles in the ground state, $N_0$, and total
particle number, $N$) are given by~\cite{Bagnato1991a,Dalfovo1999a}:
\begin{equation}
k_B T_c = \hbar \omega \left( \frac{N}{\zeta(2)} \right)^{1/2}\ ,\hspace{0.8cm}
\frac{N_0}{N}=1-\left( \frac{T}{T_c} \right)^2\ ,
\label{N_N_eqn}
\label{T_c_eqn}
\end{equation}

where $\zeta(s)$ is the {\it zeta}-function, defined as
$\zeta(s)=\sum_{n=1}^{\infty}1/n^s$.
In the 3D case these dependencies are:
\begin{equation}
k_B T_c = \hbar \omega \left( \frac{N}{\zeta(3)} \right)^{1/3}\ ,
\hspace{0.8cm}\frac{N_0}{N}=1-\left( \frac{T}{T_c} \right)^3\ .
\label{N_N_eqn3D}
\end{equation}

In the experiment we create an array of 2D condensates
by superimposing a far-detuned, standing laser wave to the long
(horizontal) axis of our static magnetic trap.
While the magnetic trapping potential is a 3D potential which confines the 
atoms to an overall cigarshaped distribution,  
in a 1D optical lattice the atoms are confined to 
2D planes.
Therefore, by increasing the depth of the nodal planes of the optical lattice
superposed to a 3D potential it is possible to follow the transition from
a 3D BEC to an array of 2D degenerate atomic clouds 
confined radially by the magnetic
potential and assorted in the axial direction like disks in a shelf.

Due to the magnetic trapping potential, the central lattice wells are
populated with a higher number of atoms, 
which -- according to Eqn.~\ref{T_c_eqn} --
leads to a higher critical temperature for the central clouds than for the
clouds in the wings of the overall density distribution.
BECs form first in the central 2D disks,
lowering the temperature leads to BEC formation at more and more lattice 
sites. 
The investigation of this successive formation of BECs
is one of the main purposes of the experiments presented in this paper.
 
\begin{figure}
\centerline{\includegraphics[scale=0.5]{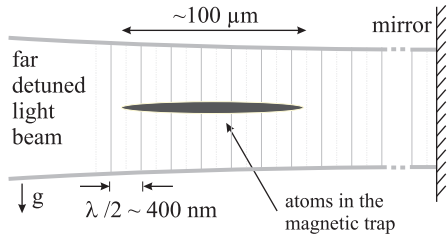}}
\caption {\small
Schematic setup of the experiment.  A blue-detuned standing-wave is superposed
to the static magnetic trap.}
\label{schema_fig}
\end{figure}

In the purely magnetic trap
(fundamental trapping frequencies 
$\omega_x=2\pi\times 8.7$\,Hz and $\omega_{\perp}=2\pi\times90\,$Hz 
along the axial and radial directions, respectively)
 we reach the phase transition to BEC at a 
temperature of $T\approx 240\,$nK with a number of $^{87}$Rb atoms of 
$N\approx  2\times 10^6$ in the quantum state F=1,\,m$_F$=\,$-1$.

The 1D optical lattice 
is created
by a retroreflected laser beam superposed 
to the long axis  of the magnetic 
trap (see Fig.~\ref{schema_fig}).
The beam is  created by a stabilized diode laser, 
with a frequency detuning of $\Delta=2\pi\,150\,$GHz
with respect to the 
D1-line of the Rubidium atoms (wavelength $\lambda = 795$\,nm, 
wavenumber $k=2\pi/\lambda$).
The dipole potential experienced by the atoms has the form
$U(\vec{r}\,)=U_0\cos^2 kx$,
the depth of the standing-wave dipole-potential wells 
amounts up to $U_0\approx 5\,E_{rec}$.
Here, $E_{rec}$ is the recoil energy of an atom in the optical lattice, 
$E_{rec}=(\hbar k)^2/2m$ with $m$ being the atomic mass.
With this detuning and intensity of the light the spontaneous-scattering rate $R$ of 
photons from the optical lattice is of the order $R=10\,$Hz.

In the experiments presented here
Bose-Einstein condensates in the combined magnetic trap and optical 
lattice are prepared by the following procedure:
First we cool the atomic cloud in the purely magnetic trap
by RF-induced evaporative cooling. At a time $t_l=50\,$ms before the end of the
RF-ramp we superpose the optical lattice ot the trapping potential
and continue to evaporatively cool the ensemble. 
In order to assure that the state reached by the atoms after the end of the RF-ramp
is a steady state of the system we use the fact that the periodically
modulated density distribution of the BECs in real space corresponds to 
a comb of equally spaced peaks in momentum space.
In a time-of-flight measurement we check that the fraction of atoms in the
different momentum components of the ground state does not depend on $t_l$ but only
on the depth of the dipole-potential wells~\cite{Cataliotti2001,note_steadys}.

\begin{figure}
\centerline{\includegraphics[scale=0.5]{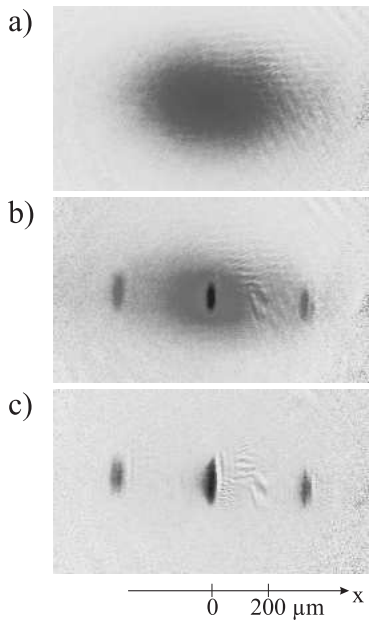}}
\caption {\small
Absorption images of a thermal cloud (a), 
a mixed cloud (b)~\cite{note_phase}, and a 
pure BEC (c), expanded for 26.5\,ms from the combined magnetic
trap and optical lattice.
The corresponding temperatures and atom numbers are 
$T\approx 210\,$nK and $N\approx 4\times 10^5$ (a), 
$T\approx 110\,$nK and $N\approx 1.5\times 10^5$ (b), 
$T < 50\,$nK and $N\approx 1.5\times 10^4$ (c).
}
\label{absbild}
\end{figure}

For reaching the quasi 2D regime, the motion of the particles 
has to be effectively ``frozen'' in the direction of the optical lattice
beam~\cite{Petrov2000a}, i.e., the fundamental frequency in a single 
lattice site, $\omega_l$ has to fulfill $\hbar \omega_l\gg k_B T$.
For our experimental parameters of $T<200\,$nK and 
$\omega_l\approx 2\pi 14\,$kHz (for $U_0\approx 4\,E_{rec}$) this relation
is well satisfied. 
Nevertheless, due to the small width of the barriers 
atoms can tunnel between the lattice sites.
The low energy of thermal atoms allows them to tunnel only over a few
sites during the duration of the experiment. 
Therefore we expect only minor changes of the thermodynamic properties due 
to such processes.
In contrast, tunnelling of ground state atoms is greatly enhanced because 
the ground state is macroscopically occupied.
As a result, the BECs at the optical lattice sites form a phase
coherent ensemble giving rise to the interference
pattern in the expansion. 
The fundamental difference in tunnelling behaviour of the thermal cloud
and the macroscopically occupied quantum fluid is also experimentally seen by
applying an external potential to a mixed cloud in a similar setup which
forces only the ground state fraction to move while the thermal distribution of atoms
sticks to its initial position~\cite{Cataliotti2001}.

\begin{figure}
\centerline{\includegraphics[scale=0.5]{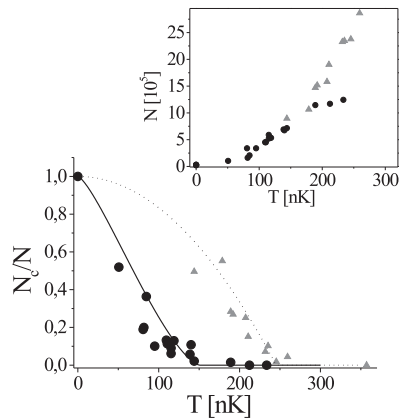}}
\caption {\small
Ground state occupation vs. temperature in the combined trap with
$V_0\approx 4\,E_{rec}$ (circles), and in the 3D purely magnetic trap (triangles).
The solid (resp. dotted) line gives the expected dependence for the combined
(resp. purely magnetic) trap, according to Eqn.~\ref{2Darray_occu} 
(resp. Eqn.~\ref{N_N_eqn3D}).
The inset shows the dependence of the total atom number on the temperature
reached by forced evaporative cooling. Also here, circles correspond to
ensembles in the combined trap and triangles to ensembles in the purely
magnetic trap.
}
\label{Nc_N_fig}
\end{figure}

In order to measure the effects of the dimensionality 
on $T_c$ and on $N_0/N (T)$ we have
varied the final temperature of the atomic clouds
and recorded the atomic density distributions by absorption imaging. 
Figure~\ref{absbild} shows  absorption images of ensembles
at different temperatures,
expanded from a combined trap with a lattice-potential height
of $U_0\approx 4\,E_{rec}$.
Due to their coherence, the interference pattern of the expanding array
of BECs 
appears in three
spatially separated peaks (in Fig.~\ref{absbild}\,b,c)~\cite{Pedri2001}.
In order to determine the temperature of the gas we fit a  
2D bimodal distribution (Gauss + Thomas-Fermi) to the central part of  the 
absorption image (zero order of the diffraction pattern of the BEC array).
For the calculation of all temperatures given in this paper we use 
the Gaussian width in the radial direction where we expect the
expansion not to be affected by the presence of the optical lattice.

By integrating over the atomic density distribution we 
obtain the number of atoms in the ground state, $N_0$, 
and in the thermal cloud, $N_{th}=N-N_0$.
Figure~\ref{Nc_N_fig} shows  the ground state
fraction $N_0/N$, in dependence of the temperature of the ensemble.
The corresponding total atom numbers are shown in the inset
of this figure~\cite{note_atomnumber}. 
In the case of the 3D potential of the pure magnetic trap 
(triangles in Fig.~\ref{Nc_N_fig}) 
this ratio reproduces the shape expected from Eqn.~\ref{N_N_eqn3D}
(dotted line, using a linear fit to the measured function $N(T)$).
The shape of the curve for ensembles produced in the
combined trap is  much smoother around the transition temperature and mixed
clouds with a relatively small condensate fraction exist in a broad
temperature range well below $T_c$.
It is also seen from Fig.~\ref{Nc_N_fig} that in the presence
of the optical lattice the transition temperature is reduced by a factor
 of $\sim 2$.

The shape of the curve $N_0/N(T)$ and the change in transition temperature 
can be qualitatively understood with
the following simplifying picture:
We assume that, before applying the optical lattice,
 the envelope of the linear atomic density of the gas in $x$-direction is
given by a Boltzmann distribution, 
$n(x)=\alpha \exp(-x^2/2a_{th}^2)$, with $\alpha = N(T)/(\sqrt{2\pi}a_{th})$ and
$a_{th}=\sqrt{kT/m\omega_x^2}$, where the atomnumber $N$ depends on temperature.
At a given temperature $T$, a single lattice site, $i$, is populated by $N_i$
atoms, $N_i=n_i \cdot\lambda /2$ ($n_i$ is the linear atomic density at  site $i$, 
$\lambda/2$ is  the width of a lattice site).
According to Eqn.~\ref{T_c_eqn} there exists a different critical temperature
$T_{ci}$ for each lattice site, and in all sites with $T_{ci}>T$ the number 
of ground state atoms is $N_{0i}$. 
Therefore the sum of particles in the
ground state is given by

\begin{equation}
N_0 (T,N)=\sum_i N_{0i}= \sum_i N_i \left( 1-\left(T \over T_{ci}\right)^2 \right) \ ,
\label{2Darray_occu}
\end{equation}

where the summation is performed over all lattice sites with $T_{ci}>T$.
This sum of the ground state occupations of the 2D BECs
is shown in the solid line in Fig.~\ref{Nc_N_fig},
where the total atom number N(T) is given by a linear fit to the
measured values.
The good agreement of the shapes of the curves indicates that
the basic principle of the phase transition to the array of 2D BECs is
accounted for by the simple model.
A more sophisticated theoretical modelling of the problem should
include interactions and effects like changes of the scattering length,
or the effect of tunnelling through the periodic lattice potentials
on the density of states and on thermodynamic properties.

In conclusion, we have experimentally 
investigated the Bose-Einstein 
phase-transition of a weakly interacting
atomic gas confined to a periodically modulated potential.
We have 
observed a change of the transition temperature and of the 
ground state occupation of the gas,
indicating that we have reached 
a regime of virtually independent formation of 2D BECs at
the single lattice sites.
As an important result we have found that in order to 
attain a high ground state occupation (``pure BEC'')
of a bosonic ensemble in an optical lattice, 
one has to cool to much lower temperatures than necessary
for a pure BEC in the corresponding magnetic trap.
The understanding of the thermodynamic properties of a BEC
in a periodic potential is crucial for future applications
of these system, e.g., in atom optics or quantum computing.
We plan to further investigate experimental signatures of 
the 2D regime
like changes in the dynamical BEC properties~\cite{Stringari_priv} 
in  future studies.

We acknowledge  discussions with
F.~Ferlaino,
G.~Ferrari, 
M.~Modugno,
L.~Pitaev\-skii, 
G.\,V. Shlyapnikov, and
S.~Stringari,  
and support by  the EU under Contracts
HPRI-CT 1999-00111 \& HPRN-CT-2000-00125,
and by MURST through  the PRIN\,1999 \& 2000  Initiatives.

\setlength{\baselineskip}{11pt plus 2pt minus 2pt}


\begin{thebibliography}{10}
\bibitem{note_steadys}
Strictly, this criterion is valid for reaching the groundstate within the
  single wells. A further indication that we reach also the thermodynamic
  steady state of the system is given by the fact that for similar starting
  conditions, very differing momentum distributions in the cloud are observed,
  ranging from thermal clouds to pure BECs, dependent on temperature (see
  Fig.~\ref{absbild}).
\bibitem{note_atomnumber}
We have calibrated the measurement of the atomic density using Eqn.~\ref{N_N_eqn3D}.
The observed reduction of $N$ is compatible with the expected loss due to light
scattering from the optical lattice beams; heating effects are neglected on the 
time scales of our experiments.
\end{thebibliography}
\end{document}